\documentstyle[12pt]{article}

\newcommand{\AmS}{{\protect\the\textfont2
  A\kern-.1667em\lower.5ex\hbox{M}\kern-.125emS}}

\newcommand{\half}{\mbox{\small$\frac{1}{2}$}}

\newcommand{\calH}{{\cal H}}
\newcommand{\be}{\begin{equation}}
\newcommand{\ee}{\end{equation}}
\newcommand{\bea}{\begin{eqnarray}}
\newcommand{\eea}{\end{eqnarray}}

\newcommand{\ph}{\phi}
\def\rellow#1#2{\mathrel{\mathop{\kern 0pt #1}\limits_{#2}}}

\newcommand{\Um}[1]{U_{#1,\mu}}

\newcommand{\Ut}[1]{U_{#1,\tau}}
\newcommand{\Tr}{\mbox{Tr}}

\newcommand{\vn}{\vec n}
\newcommand{\vs}{\vec \sigma}

\newcommand{\Lamt}{ \Lambda_t^{\tau} }

\newcommand{\delH}{\Delta\calH}

\title{\vbox{\vspace{.1mm}}
       Theoretical Analysis of Acceptance Rates in
       Multigrid Monte Carlo
       \footnote{Talk delivered by M.\ Grabenstein
       at the International Symposium on Lattice Field Theory,
       15-19 September 1992, Amsterdam} }

\author{ \vbox{\vspace{7mm}}
   {\bf Martin Grabenstein$^{1}$ and Klaus Pinn$^{2}$} \\[6mm]
$^1\,$II.\ Institut f\"ur Theoretische Physik, Universit\"at Hamburg,
      \\
      Luruper Chaussee 149, D-2000 Hamburg, Germany
      \\
      {\tt \small e-mail i02gra@dhhdesy3}
      \\[4mm]
$^2\,$Institut f\"ur Theoretische Physik I, Universit\"at M\"unster,
      \\
      Wilhelm-Klemm-Str.\ 9, D-4400 M\"unster, Germany
      \\
      {\tt \small e-mail pinn@yukawa.uni-muenster.de}     \\[4mm] }
\date{October 1992}

\begin{document}
\maketitle

\vspace{1.5cm}
\begin{abstract}
\vspace{5mm}
We analyze the kinematics of multigrid Monte Carlo
algorithms by investigating acceptance rates for nonlocal
Metropolis updates.
With the help of a simple criterion we can decide whether or not
a multigrid algorithm will have a chance to overcome critial slowing
down for a given model.
Our method is introduced in the context of spin models.
A multigrid Monte Carlo procedure for nonabelian lattice gauge
theory is described, and its kinematics is analyzed in detail.
\end{abstract}

\thispagestyle{empty}

\newpage

\section{Introduction}

Multigrid Monte Carlo algorithms were introduced to overcome
critical slowing down (CSD) \cite{multigrid}.
For some interacting models,
numerical experiments showed that the dynamical critical
exponent $z$ was indeed substantially reduced \cite{works}.
For other models, still $z \approx 2$ was found \cite{fails}.

Therefore, an improved theoretical understanding of
these algorithms is desirable.
We study the kinematics of multigrid Monte Carlo
by investigating the scale dependence of acceptance rates
for nonlocal Metropolis updates.
A simple criterion for the possible success of a multigrid
algorithm is given.
Our analysis is also useful for the design of new
multigrid algorithms for nonabelian gauge fields.

\section{Multigrid Monte Carlo}

\noindent
First we consider models with partition functions
\be
Z= \int \prod_{x \in \Lambda_0} d\ph_x \, \exp(-\calH(\ph)) \,
\ee
on cubic $d$-dimensional lattices $\Lambda_0$.
We shall use dimensionless spin variables.
Nonlocal Monte Carlo updates are defined as follows:
Divide the fundamental lattice $\Lambda_0$ in
cubic blocks of size $l^d$ (e.g. $l = 2$).
This defines a block lattice
$\Lambda_1$. By iterating this procedure one gets a
hierarchy of block lattices $\Lambda_0, \Lambda_1, \dots, \Lambda_K$.
We denote block lattice points in $\Lambda_k$ by $x'$.
Block spins $\Phi_{x'}$ are defined on block lattices
$\Lambda_k$. They are averages of the fundamental field $\ph$
over blocks of side length $L_B=l^k$:
\be
\Phi_{x'} = L^{(d-2)/2}_B \, L^{-d}_B  \sum_{x \in x'} \ph_x \, .
\ee
A nonlocal change of the configuration $\ph$ consists
of a shift
\be\label{shift}
\ph_x \rightarrow \ph_x + s \, \psi_x,
\ee
where $s$ is a real parameter.
The shape of the nonlocal change is determined by the
``coarse-to-fine interpolation kernel''
$\psi$ that obeys the constraint
\be\label{normpsi}
L^{-d}_B \sum_{x \in x'} \psi_x = L^{(2-d)/2}_B \delta_{x',x_o'} \, .
\ee
Note that the effect of (\ref{shift}) on $\Lambda_k$ is
$ \Phi_{x'} \rightarrow \Phi_{x'} + s $
for $x'=x_o'$, whereas
$\Phi_{x'}$ remains unchanged on the other blocks.
The simplest $\psi$ is a piecewise constant
kernel: $\psi_x = L^{(2-d)/2}_B$ if $x \in x_o'$, and $0$ else.
One can also use smooth kernels that avoid
large energy costs from the block boundaries.

\noindent
The $s$-dependent Metropolis acceptance rate for such
proposals is given by
\be\label{omega}
\Omega(s) = \bigl<
\min \lbrack 1 , \exp( - \Delta \calH) \rbrack \bigr> \, ,
\ee
where
$ \Delta \calH = \calH(\ph + s\psi)-\calH(\ph)$.
The starting point of our acceptance analysis
is the approximation formula
\cite{hybrid,letter,long}
\be\label{formula}
\Omega(s) \approx
\mbox{erfc} \left( \half\sqrt{h_1} \right) \, .
\ee
Here, $h_1 = \langle \delH \rangle$ denotes the average change in the
fundamental Hamiltonian.
Generally, the formula yields precise estimates that are
confirmed by the acceptance rates directly measured in Monte Carlo
simulations \cite{letter,long}.
We will use (\ref{formula}) to predict the
acceptance rate $\Omega(s)$ for interacting models.
Let us first discuss free massless field theory with action
$\calH(\ph)= \half(\ph,-\Delta \ph)$.
Here, we obtain the exact result
\be
\Omega(s) = \mbox{erfc}(\sqrt{\alpha/8} \vert s \vert) \, ,
\ee
with $\alpha=(\psi,-\Delta\psi)$.
In $d$ dimensions one finds
$ \alpha = 2 d L_B$ for piecewise constant kernels,
and, for smooth kernels,
$\alpha \rightarrow \mbox{const} $ if $L_B >\!\!> 1$.
(For a systematic study of different kernels see ref. \cite{long}.)
As a consequence, in massless free field theory, to maintain
a constant acceptance rate (of, say, 50 percent) the
amplitudes $s$ have to be scaled down like $L^{-1/2}_B$ for piecewise
constant kernels, whereas for smooth kernels the acceptance
rates do not depend on the block size.
(At least for free field
theory, the disadvantage of the piecewise constant kernels can
be compensated for by using a W-cycle instead of a V-cycle.
Smooth kernels can be used only in V-cycle algorithms.)

\section{Spin Models}

Now we discuss multigrid procedures for spin models.
The kinematical analysis is a comparison of
the scale dependence of acceptance rates for interacting models
with the behavior in free field theory,
where CSD is known to be eliminated by a
multigrid algorithm.

\noindent
As an example we consider the 2-dimensional Sine Gordon model
defined by the Hamiltonian
\be
\calH(\ph) = \frac1{2\beta} ( \ph, -\Delta \ph)
- \zeta \sum_x \cos \ph_x  \, .
\ee
The model undergoes a (Kosterlitz-Thouless) phase transition
at $\beta_c$, and $\beta_c \rightarrow 8 \pi$ for $\zeta \rightarrow 0$.
In the massless phase ($\beta > \beta_c$),
the long wavelength excitations are spin waves.
Since multigrid is an efficient method to accelerate spin waves in
free massless field theory, one might expect
that multigrid should be also the right method to fight
CSD in the massless
phase of the Sine Gordon model.
But this expectation is wrong:
For $h_1$ we find the expression
\be\label{h1h2}
h_1 = \frac{\alpha}{2\beta} s^2
+ \zeta C \sum_x \lbrack  1 - \cos(s\psi_x) \rbrack \, ,
\ee
with $C = \langle \cos \ph_x \rangle$.
Expanding for small $s$, we find that
the second term in (\ref{h1h2}) behaves like
$\sim s^2 \sum_x\psi_x^2$, like a ``mass'' term in the kernel $\psi$.
Since $\sum_x\psi_x^2$ is proportional
to the block volume $L^2_B$ for piecewise constant and
for smooth kernels, one therefore has to face a dramatic
decrease of acceptance when the blocks become large.
A constant acceptance rate can only be
achieved when the proposed steps $s$ are scaled down like $L^{-1}_B$.
It is therefore unlikely that any multigrid algorithm
 - based on nonlocal updates of the type discussed here -
will be successful for this model.

Analyzing multigrid algorithms, we found two classes of models
\cite{letter,long}.
For 2d Sine Gordon and $\phi^4_d$ theory,
$s$ has to be rescaled like $L^{-1}_B$ for piecewise
constant and for smooth kernels, whereas for massless free field theory,
the 2d $XY$ model, the 2d $O(N)$ nonlinear $\sigma$-model and
$U(1)$ lattice gauge theory, one can achieve $L_B$-independent
acceptance rates choosing smooth kernels.

For almost all models of the second class, at least a
substantial reduction of CSD could be achieved \cite{works}.
An exception is the 2d $XY$ model in the vortex phase.
There $z \approx 1.4$ was found \cite{XY}.
This shows that good acceptance rates alone are not sufficient to
overcome CSD.

The results of the analysis are consistent with the following rule:
Sufficiently high acceptance rates for a complete elimination of
CSD can only be expected if $\, h_1 = \langle \calH(\ph+s \psi) -
\calH(\ph) \rangle $ contains no algorithmic ``mass'' term
$\sim s ^2 \sum_x \psi^2_x$.

\section{Nonabelian Gauge Fields}

\label{SECgauge}

We study 4-dimensional $SU(2)$
lattice gauge theory with standard Wilson action
\be \label{wilson-action}
{\cal H}(U)\,=\,\beta\sum_{\cal P} \bigl[ 1 -
\half \, \Tr \, U_{\cal P} \bigr] \ \ .
\ee
The $U_{\cal P}$ are the usual path ordered
products around plaquettes ${\cal P}$.

\subsection{Covariant nonlocal update proposal}
\label{SUBSECcov}

We want to update only link variables $\Ut{x}$
pointing in a selected direction $\tau$.
Then only $\Ut{x}$ variables
in the same 3-dimensional slice $ \Lambda_t^{\tau} =
\left\{ x \in \Lambda_0 \, \vert \, x_{\tau} = t \right\}$
are coupled
(when all other link variables $\Um{x}$ with
$\mu \neq \tau$  are kept fixed).
Regard the $\Ut{x}$ variables as spins.
We then perform updates for a
3-dimensional spin model in a disordered background field given
by the fixed link variables in the directions $\mu \neq
\tau$.
Now build 3-dimensional blocks $x_o'$ of size $L_B^3$ that are
contained in the slice $\Lamt$.
Nonlocal update proposals will be global rotations of all link
variables $\Ut{x}$ attached to sites $x$ inside the block $x_o'$ by
multiplying them with an $SU(2)$-matrix.

Such global rotations only make sense
if we have a certain smoothness of gauge fields inside the slice
$\Lamt$. We define a gauge transformation $g$ by the
Coulomb gauge condition
\be
G_{C}(U,g) =\!\! \sum_{(x,x+\hat\mu) \in \Lamt}\!\!\!\! \Tr \bigl(
g_x \Um{x} g_{x+\hat\mu}^*\bigr)
 \,\stackrel{\mbox{!}}{=} \, \mbox{max.}
\ee
If we actually performed this gauge transformation
$\Um{x}^g = g_x \Um{x} g_{x+\hat\mu}^* $
we would end up with the gauge fields fixed to the
Coulomb gauge.
However, we will use the transformation $g$ only to
achieve a smooth rotation of the $\Ut{x}$ variables inside the block.
This smooth rotation is defined as follows:

Propose new link variables $\Ut{x}'$ by
\be
\Ut{x} \rightarrow \Ut{x}' = g_x^* R_x g_x \, \Ut{x} \, ,
\ee
with
\be
R_x(\vn,s) = \cos( s \psi_x /2 )
 + i \sin( s \psi_x /2) \,
\vn \!\cdot\! \vs \, .
\ee
$s$ is a uniformly distributed random number from the
interval $[-\varepsilon,\varepsilon]$, $\vn$ is a 3-dimensional
random unit vector,
and $\vs$ are the Pauli matrices.
$\psi$ again denotes a kernel normalized as in (\ref{normpsi}).
Finally calculate the associated change of the Hamiltonian
$\Delta \calH$ and accept the proposed link variables with probability
$ \min \lbrack 1,\exp(-\Delta \calH) \rbrack $.

One can show that this is a valid algorithm \cite{long}.
For the detailled balance condition it is crucial to
ensure that the updates are reversible.
We have to get the same $g$ before and after the move $\Ut{x}
\rightarrow \Ut{x}'$. This is fulfilled here, since only
link variables $\Um{x}$ with $\mu \neq \tau$ enter in the Coulomb
gauge functional.

Note that we do not have to extremize the gauge functional perfectly.
If we always use the same algorithm (e.g.\ a given number of relaxation
sweeps starting from $g=1$), we will always get the same $g$ and the
nonlocal update is reversible.

\begin{table*}[hbt]
\begin{center}
\caption{Comparison of $m_D$ with physical masses}
\label{tab1}
\begin{tabular}{|ccccccc|}
\hline
 lattice size & $\beta$    & $m_D$     & $\sqrt{\kappa}$  &
 $m_{0^+}$    & $m_D/\sqrt{\kappa}$  & $m_D/m_{0^+}$ \\
\hline
$16^4$ & $2.4$ & $0.4955(2)$ & $0.258(2)$ & $0.94(3)$ &
$1.92$ & $0.53$ \\
$20^4$ & $2.6$ & $0.4650(2)$ & $0.125(4)$ & $0.52(3)$ &
$3.72$ & $0.89$ \\
\hline
\multicolumn{7}{c}{Estimates for physical masses are taken from
                          ref.\ \cite{michael}}
\end{tabular}
\end{center}
\end{table*}
\subsection{Acceptance analysis for $SU(2)$}
\label{SUBSECaca}

Now we are going to check whether our proposal has a chance
to overcome CSD.
For the quantity $h_1 = \langle \delH \rangle$ that enters in
the approximation formula (\ref{formula}) for the acceptance rates,
we get (using piecewise constant $\psi$)
\bea \label{h1_constant}
h_1 &=&
3 A \, (L_B -1) L_B^2 \, \sin^2(s L_B^{-1/2}/2)
                                      \nonumber  \\
    &+&
     6 P \, L_B^2 \bigl[ 1- \cos(s L_B^{-1/2}/2) \bigr] \, ,
\eea
with
\be A= - \frac{\beta}2
      \bigl\langle \Tr \bigl((\vn \!\cdot\!  \vs \, \Um{x}^g \,
\vn\!\cdot\!\vs - \Um{x}^g) H_{x,\mu}^{g\, *} \bigr) \bigr\rangle\, ,
\ee
$H_{x,\mu}^* = \Ut{x+\hat\mu} \Um{x+\hat\tau}^* \Ut{x}^*$,
and $P = \beta/2 \langle \Tr U_{\cal P} \rangle$.
If we expand $h_1$ for small $s$, we see that the second term (coming
from plaquettes that have one link in common with the block)
is linear in $L_B$, the usual surface energy with piecewise constant
kernels.
But the first term (coming from pla\-quettes that are entirely inside
the block) is again a ``mass'' term:
it  behaves like $\sim A s^2 L_B^2$.

Let us discuss the origin of this term in more detail:
In the weak coupling limit $\beta \rightarrow \infty$ we have only
pure gauges, i.e. $\Um{x}^g \rightarrow 1$ (with the
gauge transformation $g$ defined as above).
Then the difference of the two terms contributing to $A$ vanishes,
and $A \rightarrow 0$.
In this limit the unwanted mass term vanishes, and the acceptance rates
behave just like in massless free field theory.

For finite $\beta$, however, $\Um{x}^g \neq 1$ and therefore
$A > 0$ because of the disorder of the gauge fields inside the block.
Motivated by this, we identify the square root of $A$ with a ``disorder
mass'' $m_D = \sqrt{A}$.
This disorder mass term will dominate the average energy change $h_1$
for large $L_B$
\be
h_1 = \langle \delH \rangle \rellow{\sim}{L_B >\!\!> 1}
m_D^2 s^2 L_B^2   \, .
\ee
Close to criticality one might hope that
$m_D$ scaled with physical masses in the theory, say,
the square root of the string tension $\kappa$
or the lowest glue ball mass $m_{0^+}$.
Then, the fluctuations created by the algorithm would behave
similar to the physical fluctuations, and we would expect
the algorithm to behave well.
\noindent
But we have to face the fact that
$m_D$ will be dominated by the local disorder and
decreases much slower than the physical masses in the
system. This will be examined now.

\subsection{Monte Carlo study of $m_D$}

We computed $m_D$ for several values of $\beta$.
To maximize $G_C$ we used 50 Gauss-Seidel relaxation
sweeps. Tests showed that
increasing the number of relaxation sweeps beyond 50 hardly lowered
$m_D$ any further.

In table 1 we display the ratios of the disorder mass $m_D$
with physical masses.
The results show that the
disorder mass is nearly independent of $\beta$ in the range studied,
whereas the physical masses decrease by roughly a factor of two.
In order to keep constant acceptance rate for increasing
block size $L_B$,
one again has to rescale $s \sim 1/L_B$.
Updating on large blocks gets essentially ineffective,
and we have to expect CSD for this algorithm.

\subsection{Implementation and Test}
We implemented and tested the algorithm with 3-dimensional
blocks using piecewise constant kernels and a W-cycle
as described in ref. \cite{long}.
We did not observe any substantial speed up compared to a local heat
bath algorithm (not even a constant factor). This is in agreement
with our prediction that updates on larger blocks are
not efficient.

\section*{Conclusions}

The kinematical mechanism that leads to a
failure of multigrid algorithms is well described by our analysis.
We hope that a better understanding of this problem
can lead to improved multigrid algorithms that can overcome kinematical
obstructions stemming from an algorithmic ``mass'' term.

\section*{Acknowledgments}
\noindent
The numerical computations were
performed on the CRAY Y-MP of the HLRZ in J\"ulich.
M.G.\ would like to thank the Deutsche Forschungsgemeinschaft
for financial support.


\end{document}